\documentstyle[epsf,sprocl]{article}

\begin{document}

\title{ON ARCS AND $\Omega$}

\author{M. BARTELMANN, A. HUSS, J.M. COLBERG}

\address{Max-Planck-Institut f\"ur Astrophysik, P.O. Box 1523,
D--85740 Garching}

\author{A. JENKINS, F.R. PEARCE}

\address{Physics Dept., University of Durham, Durham DH1 3LE, UK}

\maketitle

\abstracts{The gravitational lens effect of galaxy clusters can
produce large arcs from source galaxies in their background. Typical
source redshifts of $\sim1$ require clusters at $z\sim0.3$ for arcs to
form efficiently. Given the cluster abundance at the present epoch,
the fewer clusters exist at $z\sim0.3$ the higher $\Omega_0$ is,
because the formation epoch of galaxy clusters strongly depends on
$\Omega_0$. In addition, at fixed $\Omega_0$, clusters are less
concentrated, and hence less efficient lenses, when the cosmological
constant is positive, $\Omega_\Lambda>0$. Numerical cluster
simulations show that the expected number of arcs on the sky is indeed
a sensitive function of $\Omega_0$ and $\Omega_\Lambda$. The numerical
results are compatible with the statistics of observed arcs only in a
universe with low matter density, $\Omega_0\sim0.3$, and zero
cosmological constant. Other models fail by one or two orders of
magnitude, rendering arc statistics a sensitive probe for cosmological
parameters.}

\section{Introduction}

Given the average mass of a rich galaxy cluster, $M_{\rm c}$, and the
spatial number density of such clusters in our neighbourhood, $n_{\rm
c}$, we can find the fraction of cosmic material contained in rich
clusters today,
\begin{equation}
  F_{\rm c}' = \frac{M_{\rm c}n_{\rm c}}{\rho_{\rm cr}\Omega_0}
  \approx \frac{9\times10^{-3}}{\Omega_0}\;,
\label{eq:1}
\end{equation}
where $\rho_{\rm cr}$ is the critical cosmic matter density, and
$\Omega_0$ is the usual density parameter. The {\em ansatz\/} by Press
\& Schechter (1974) asserts that the fraction of cosmic material
accumulated by clusters at redshift $z$ is
\begin{equation}
  F_{\rm c}(z) = \frac{1}{2}\,{\rm erfc}\left(
  \frac{\delta_{\rm c}}{\sqrt{2}\sigma_{\rm c}{\cal D}(z)}\right)\;,
\label{eq:2}
\end{equation}
where $\sigma_{\rm c}$ is the {\em rms\/} density fluctuation on the
linear cluster scale today, and ${\cal D}(z)$ is the growth factor for
cosmic structures, normalized to ${\cal D}(0)=1$. $\delta_{\rm c}$ is
the linear density contrast of a spherical top-hat perturbation at
collapse time, and ${\rm erfc}(x)$ is the complementary error
function. Demanding $F_{\rm c}(0)=F_{\rm c}'$, $\sigma_{\rm c}$ is
fixed to the local cluster abundance. The evolution of $F_{\rm c}(z)$
with redshift then depends on cosmology because the growth factor
${\cal D}(z)$ does. This leads to the well-known results that (i)
clusters form late in cosmic history, and (ii) cluster formation is
significantly delayed in high-density compared to low-density
universes (e.g.\ N.\ Bahcall, these proceedings).

The ability of a galaxy cluster to act as a strong gravitational lens
(i.e., to produce arcs) depends on the geometry of the lens
system. Let $D_{\rm eff}=D_{\rm d}D_{\rm ds}D_{\rm s}^{-1}$ be the
effective lens distance, with $D_{\rm d,ds,s}$ the angular-diameter
distances between observer and lens, lens and source, and observer and
source, respectively. $D_{\rm eff}$ is a measure for the lensing
efficiency of a given mass distribution. $D_{\rm eff}$ peaks at
$z\sim0.2-0.3$ for sources at a typical redshift $z_{\rm s}\sim1$,
quite independent of cosmology.

It follows that an efficient formation of arcs requires that there be
sufficiently many clusters in place and compact enough at redshifts
$z\sim0.2-0.3$. This establishes the link between arc statistics and
cosmology. Quite obviously, the number of efficient cluster lenses per
unit redshift is estimated by
\begin{equation}
  \frac{{\rm d}N_{\rm lens}}{{\rm d}z} \propto
  F_{\rm c}(z) \times (1+z)^3 \times D_{\rm eff}^2 \times
  \left|\frac{{\rm d}V(z)}{{\rm d}z}\right|\;,
\label{eq:3}
\end{equation}
where the square on $D_{\rm eff}$ approximates the dependence of the
lensing cross section on $D_{\rm eff}$, and ${\rm d}V(z)$ is the
proper cosmic volume of a spherical shell of radius $z$ and width
${\rm d}z$. Figure~1 illustrates ${\rm d}N_{\rm lens}/{\rm d}z$ as a
function of redshift for $\Omega_0=1$ and $\Omega_0=0.3$, both for
$\Omega_\Lambda=0$. Evidently, there is a huge difference of about two
orders of magnitude, clusters in the low-density universe being much
more efficient in producing arcs than in the Einstein-de Sitter
universe. This straightforward argument leads one to expect that the
number of observed arcs could be a sensitive discriminator between
cosmological models.

\begin{figure}[ht]
  \parbox[c]{0.55\textwidth}{%
    \epsfxsize=\hsize\epsffile{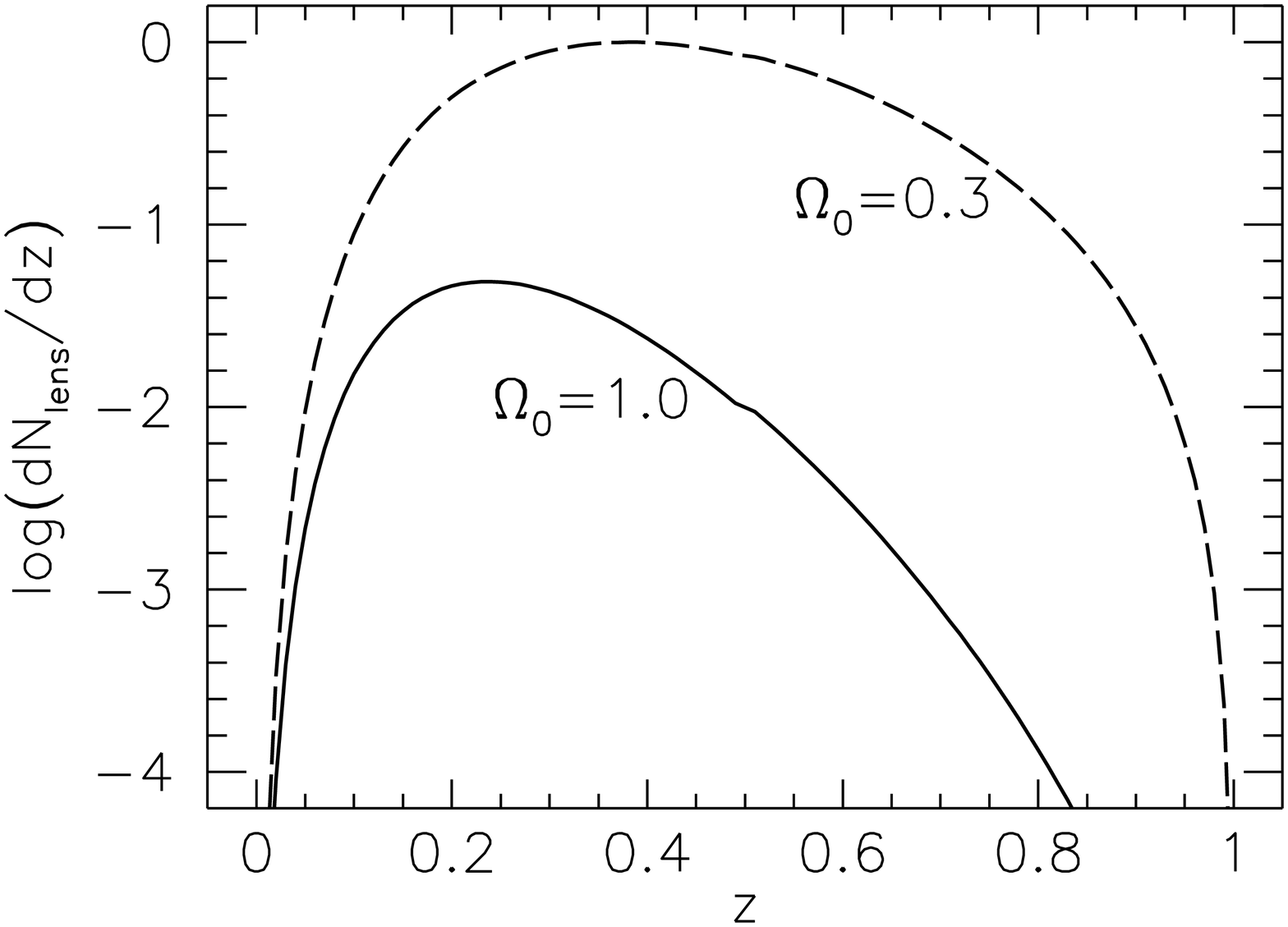}}
  \parbox[c]{0.44\textwidth}{%
    \small Fig.~1 --- Estimate of the number of lenses per unit
      redshift, ${\rm d}N_{\rm lens}/{\rm d}z$, as defined in the
      text, for high- and low-density universes. Note the logarithmic
      scale of the ordinate. The curves illustrate that the number of
      efficient lenses is expected to be lower by about two orders of
      magnitude in high- compared to the low-density model.}
\end{figure}

\section{Simulations}

We use numerical simulations to quantify the expected effect more
precisely. Clusters are taken from four different cosmological
simulations, the parameters of which are summarized in Tab.~1. The
models indexed by 1 were kindly supplied by the GIF collaboration
(cf.\ S.D.M.\ White or J.M.\ Colberg, these proceedings), those
indexed by 2 were simulated with a different numerical algorithm. In
total, we use nine simulated clusters for (S,$\Lambda$,O)CDM, and five
for $\tau$CDM.

\begin{table}[ht]
  \parbox[c]{0.55\textwidth}{%
    \small
    \begin{tabular}{l|*{7}{c}}
    \hline
    Model & $\Omega_0$ & $\Omega_\Lambda$ & $h$ & $\sigma_8$ &
    $\Gamma$ \\
    \hline\hline
    SCDM1         & 1.0 & 0.0 & 0.5 & 0.60 & 0.50 \\
    $\tau$CDM1    & 1.0 & 0.0 & 0.5 & 0.60 & 0.21 \\
    $\Lambda$CDM1 & 0.3 & 0.7 & 0.7 & 0.90 & 0.21 \\
    OCDM1         & 0.3 & 0.0 & 0.7 & 0.85 & 0.21 \\
    \hline
    SCDM2         & 1.0 & 0.0 & 0.5 & 0.60 & 0.50 \\
    $\Lambda$CDM2 & 0.3 & 0.7 & 0.7 & 1.12 & 0.21 \\
    OCDM2         & 0.3 & 0.0 & 0.7 & 1.12 & 0.21 \\
    \hline
    \end{tabular}}
  \parbox[c]{0.44\textwidth}{%
    \small Tab.~1 --- Summary of the parameters used for the cluster
      simulations. $h$ is the Hubble constant in units of $100\,{\rm
      km\,s^{-1}\,Mpc^{-1}}$, $\Gamma$ is the shape parameter of the
      power spectrum, and the other parameters have their conventional
      meaning.}
\end{table}

Each cluster is studied at ten time steps between redshifts 1 and 0,
projecting it along each of the three independent spatial directions,
ending up with roughly $10^3$ lensing mass distributions. For each of
these, the arc cross section is computed numerically, mapping
elliptical sources at redshift $z_{\rm s}=1$ that are placed on an
adaptive grid tracing the caustic curves of the lenses. In total, we
classify all images of about $1.3\times10^6$ sources. This procedure
yields arc cross sections as a function of cluster redshift,
$\sigma(z)$, for the four cosmological models used. The arc optical
depth, i.e., the probability for a source to be imaged as an arc with
specified properties, is then given by a volume-weighted integral of
$\sigma(z)$ over redshift, multiplied by the cluster number density
$n_{\rm c}$ and divided by the area of the source sphere.

\section{Results}

The optical depth normalized by the cluster number density, $n_{\rm
c}^{-1}\tau$, is shown in Fig.~2 for the models
(S,$\Lambda$,O)CDM. The result for the $\tau$CDM model is almost
identical to that for the SCDM model, and is therefore omitted. In
summary, the optical depth for large arcs, i.e. such with a
length-to-width ratio $\ge10$, is highest for the OCDM model and
lowest for the SCDM model, with differences of about an order of
magnitude between each of the models.

\begin{figure}[ht]
  \parbox[c]{0.60\textwidth}{%
    \epsfxsize=\hsize\epsffile{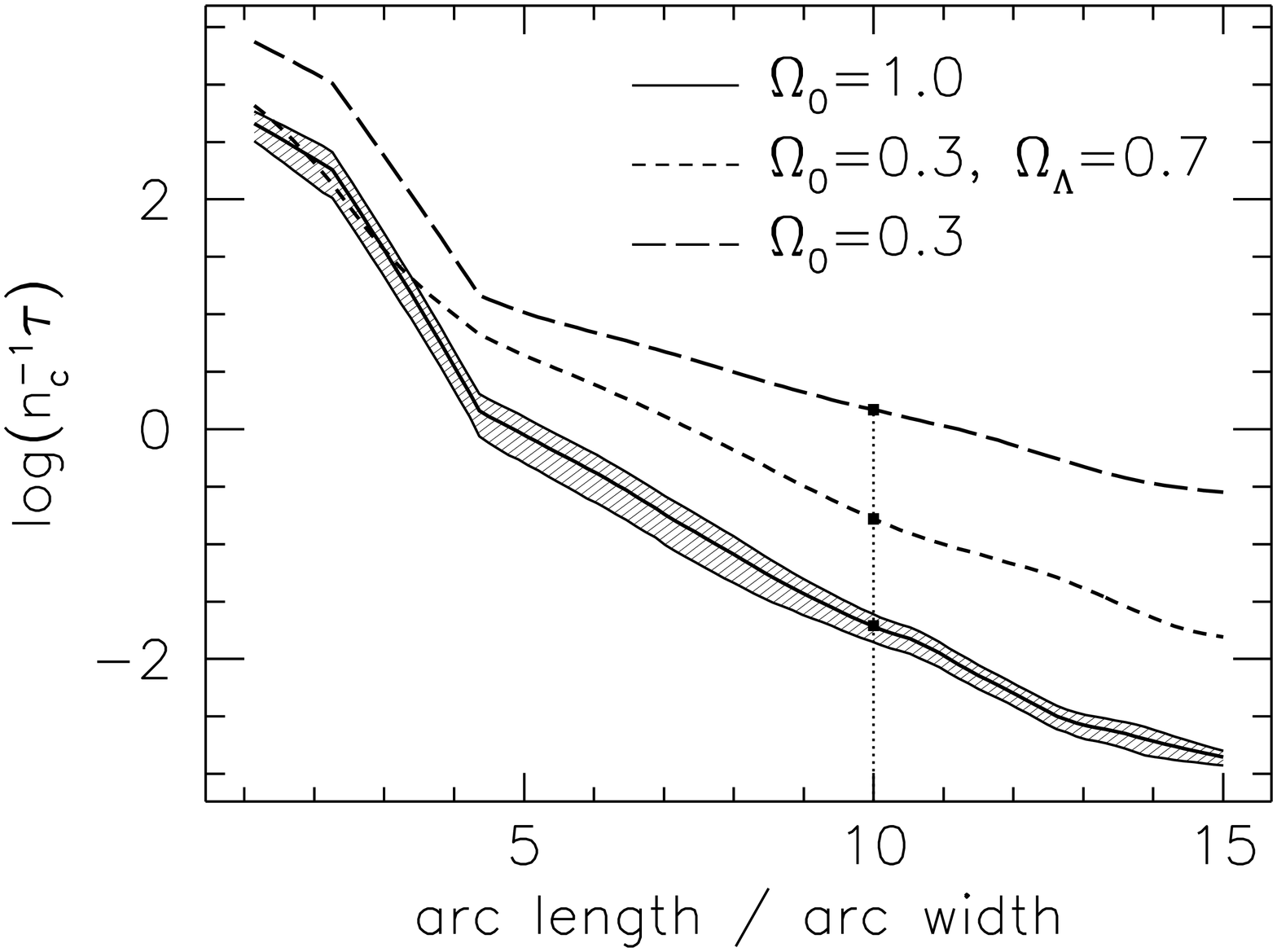}}
  \parbox[c]{0.39\textwidth}{%
    \small Fig.~2 --- Optical depth $n_{\rm c}^{-1}\tau$, as a
      function of the length-to-width ratio of the arcs. The hatched
      region centered on the SCDM curve (solid line) illustrates
      1$\sigma$ bootstrapping errors. The dots mark the optical depth
      for arcs with length-to-width ratio $\ge10$. The optical depth
      is highest for the OCDM model, and lower by one order of
      magnitude each for $\Lambda$CDM and SCDM, respectively.}
\end{figure}

Combining the optical depth with the number densities of observed
clusters and of appropriately bright sources, we find that the
number of arcs on the whole sky expected from our simulations is
\begin{equation}
 N_{\rm arcs} \sim \left\{\begin{array}{rl}
   2400 & \hbox{OCDM} \\
    280 & \hbox{$\Lambda$CDM} \\
     36 & \hbox{SCDM} \\
 \end{array}\right.\;.
\label{eq:4}
\end{equation}
The observed number of arcs, estimated from the EMSS arc survey and
extrapolated to the whole sky, falls within $1500-2300$. We therefore
conclude that {\em the only of our cosmological models for which the
expected number of arcs comes near the observed number is the open CDM
model\/}. The others fail by one or two orders of magnitude. This
result can be understood by (i) the delayed cluster formation in
high-$\Omega_0$ universes, combined with lensing efficiency, and (ii)
the higher concentration of clusters in low-density models without
$\Omega_\Lambda$ compared to such with $\Omega_\Lambda>0$, combined
with the sensitivity of lensing to compactness. For details, see
Bartelmann et al.\ (1997).

\section*{Acknowledgments}

This work was supported in part by the Sonderforschungsbereich 375 of
the Deutsche Forschungsgemeinschaft.


\section*{References}

\begin{thebibliography}{99}

\bibitem{ref:1} Bartelmann, M., Huss, A., Colberg, J.M., Jenkins, A.,
Pearce, F.R., 1997, A\&A, submitted; preprint astro-ph/9707167
\bibitem{ref:2} Press, W.H., Schechter, P.L., 1974, ApJ, 187, 425

\end{thebibliography}
\end{document}